# Polarization modulation by tunable electromagnetic metamaterial reflector/absorber

**Bo Zhu, Yijun Feng[*] Junming Zhao, Ci Huang, Zhengbin Wang, Tian Jiang**

*Department of Electronic Engineering, School of Electronic Science and Engineering,
Nanjing University, Nanjing, 210093, China*
[*]yjfeng@nju.edu.cn

**Abstract:** We propose a polarization modulation scheme of electromagnetic (EM) waves through reflection of a tunable metamaterial reflector/absorber. By constructing the metamaterial with resonant unit cells coupled by diodes, we demonstrate that the EM reflections for orthogonal polarized incident waves can be tuned independently by adjusting the bias voltages on the corresponding diodes. Owing to this feature, the reflected EM waves can be electrically controlled to a linear polarization with continuously tunable azimuth angle from $0^o$ to $90^o$ at the resonant frequency, or an elliptical polarization with tunable azimuth angle of the major axis when off the resonant frequency. The proposed property has been verified through both numerical simulations and experimental measurements at microwave band, which enables us to electrically modulate the polarization state of EM waves flexibly.

## References and links

1. Q. Chen and X. –C. Zhang, "Polarization modulation in optoelectronic generation and detection of terahertz beams," Appl. Phys. Lett. **74**, 3434-3437 (1999).
2. Y. Hirota, R. Hattori, M. Tani, and M. Hangyo, "Polarization modulation of terahertz electromagnetic radiation by four-contact photoconductive antenna," Opt. Express **14,** 4486-4493 (2006).
3. S. Betti, G. D. Marchis, and E. Iannone, "Polarization modulated direct detection optical transmission systems," Journal of Lightwave Technology **10**, 1985-1997 (1992).
4. K. Gallo J. Badoz, M. Billardon, J. C. Canit, and M. F. Russel, "Sensitive devices to determine the state and degree of polarization of a light beam using a birefringence modulator," J. Opt. **8**, 373-384 (1977).
5. R. Shimano, H. Nishimura, and T. Sato, "Frequency tunable circular polarization control of terahertz radiation," Jpn, J. Appl. Phys. **44**, L676-L678 (2005).
6. J. Hao, Y. Yuan, L. Ran, T. Jiang, J. A. Kong, C. T. Chan, and L. Zhou, "Manipulating electromagnetic wave polarizations by anisotropic metamaterials," Phys. Rev. Lett. **99**, 063908 (2007).
7. A. C. Strikwerda, K. Fan, H. Tao, D. V. Pilon, X. Zhang, and R. D. Averitt, "Comparison of birefringent electric split-ring resonator and meanderline structures as quarter-wave plates at terahertz frequencies," Opt. Express **17**, 136-149 (2009).
8. J. Y. Chin, M. Lu, and T. J. Cui, "Metamaterial polarizers by electric-field-coupled resonators," Appl. Phys. Lett. **93**, 251903 (2008).
9. A. Demetriadou and J. B. Pendry, "Extreme chirality in Swiss roll metamaterials," J. Phys.: Condens. Matter **21**, 376003 (2009).
10. Y. Ye and S. He, "$90^o$ polarization rotator using a bilayered chiral metamaterial with giant optical activity," Appl. Phys. Lett. **96**, 203501 (2010).
11. T. Q. Li, H. Liu, T. Li, S. M. Wang, F. M. Wang, R. X. Wu, P. Chen, S. N. Zhu, and X. Zhang, "Magnetic resonance hybridization and optical activity of microwave in a chiral metamaterial," App. Phys. Lett. **92**, 131111 (2008).
12. B. Zhu, Y. Feng, J. Zhao, C. Huang, and T. Jiang, "Switchable metamaterial reflector/absorber for different polarized electromagnetic waves," Appl. Phys. Lett. **97**, 051906 (2010).
13. B. Zhu, Z. Wang, Z. Yu, Q. Zhang, J. Zhao, Y. Feng, and T. Jiang, "Planar metamaterial absorber for all wave polarizations," Chin. Phys. Lett. **26**, 114102 (2009).
14. J. D. Jackson, *Classical Electrodynamics* (Wiley, New York, 1999).

15. H.-T Chen, W. J. Padilla, J. M. O. Zide, A. C. Gossard, A. J. Taylor, and R. D. Averitt, "Active terahertz metamaterial devices," Nature **444**, 597-600 (2006).
16. D. Huang, E. Poutrina, and D. R. Smith, "Analysis of the power dependent tuning of a varactor-loaded metamaterial at microwave frequencies," Appl. Phys. Lett. **96**, 104104 (2010).
17. D. A. Powell, I. V. Shadrivov, and Y. S. Kivshar, "Nonlinear electric metamaterials," Appl. Phys. Lett. **95**, 084102 (2009).

## 1. Introduction

Polarization is an important characteristic of electromagnetic (EM) waves. It has been utilized widely in many EM applications, such as in microwave communication systems, liquid crystal display, and many optical instruments. It is always desirable to manipulate polarization states flexibly. In the area of terahertz or optical spectroscopy or transmission system, polarization modulation technique is preferred to generate and detect the terahertz or optical waves so as to improve the system signal-to-noise ratio [1-3]. Conventional methods to modulate polarizations include using photoelastic modulator in infrared and mid-infrared region [4], phase control of two orthogonal polarized terahertz pulses by using a Michelson interferometer [5] or biasing four-contact photoconductive antenna [2], etc. As a result, a modulated EM wave can be generated with both left-handed and right-handed circular polarization states.

In recent years, metamaterials, as an artificial material, open the access to controlling material constitutive parameters in terms of values, distributions, anisotropy and chirality by material structure designing. Owing to their unique features, metamaterials have been applied to manipulate EM characteristics including polarizations, such as polarization manipulation through anisotropic metamaterials [6-8] and chiral metamaterials [9-11] in microwave and terahertz regimes. For polarization modulation, it is still required to have flexible control (preferable of electrical control) of polarization in transmission or reflection of the EM waves. Recently, we reported a tunable metamaterial EM wave reflector/absorber by integrating microwave diodes into the metamaterial resonant inclusions which are arranged in orthogonal directions [12]. By tuning the bias voltage on the diodes that couple the resonant unit cells, the EM reflection of the metamaterial for orthogonal polarized EM waves can be tuned continuously and independently. In this paper, we demonstrate a polarization modulation scheme at the microwave band based on the EM reflection of the tunable metamaterial reflector/absorber. Firstly, we give a detailed introduction to the metamaterial structure and its tunable reflection characteristics. Then, we demonstrate that by illuminating the metamaterial with a specific linearly polarized EM wave and tuning the bias voltage on diodes, the reflected EM wave can be linearly polarized with the polarization angle continuously controlled from $0^o$ to $90^o$ at the resonant frequency, or the reflection can be right-handed or left-handed elliptically polarized with the major axis angle controlled between $0^o$ and $90^o$ when operated off the resonant frequency. Thus, the proposed scheme allows us to electrically modulate the EM wave polarization flexibly.

## 2. Tunable metamaterial EM reflector/absorber

The tunable metamaterial EM reflector/absorber is based on the polarization-independent EM wave absorber [13]. It is composed of array of electric LC (ELC) resonators on the top layer of a dielectric substrate with a metallic sheet on the bottom layer, as shown in Fig. 1(a). The ELC structures can couple to electric component of the incident EM waves parallel to the central strip of the ELC, and provide an independent electric response. The ELC on the top layer and metallic sheet on the bottom layer form the parallel conducting structure, which couples to the magnetic components of the incident wave and supplies an independent magnetic response. Through optimizing the ELC dimensions and substrate thickness, the wave impedance of the metamaterial can be matched to that of free space and high losses can be introduced inside the resonant inclusions, leading to nearly perfect wave absorption with little wave reflection.

Single ELC is a polarization sensitive resonant structure. To respond to both EM polarizations of the incident waves equally, we combine two sets of ELCs with orthogonal orientations. The ELCs with central strips along *x* axis (*y* axis) only respond to *x* polarized (*y* polarized) incident waves. The two ELCs with the same orientation are coupled by a microwave diode as illustrated in Fig. 1(a). The resonance and absorbing strength of the metamaterial can be controlled by tuning the bias voltage applied on the diode. The diodes in different rows (row A and row B) are connected in series through the vias to two independent bias circuit networks arranged on the bottom side of the metamaterial substrate. At resonance frequency, opposite charges are accumulated on the adjacent ELC arches which are connected by diodes. When a zero bias voltage is applied on the diode, the diode is in OFF state with a large resistance in series with a diode capacitance, which enables the opposite charges to accumulate and establish a strong resonance. Whereas, when a forward bias voltage is applied on the diode, the diode is in ON state with a small resistance in series with an inductance so that the opposite charges are cancelled out and the resonance vanishes. By adjusting the bias voltages on the diodes from zero bias to forward bias (Fig. 1(a)), we are able to tune the EM absorption (reflection) of the metamaterial for an incident EM wave from high (low) level to low (high) level [12].

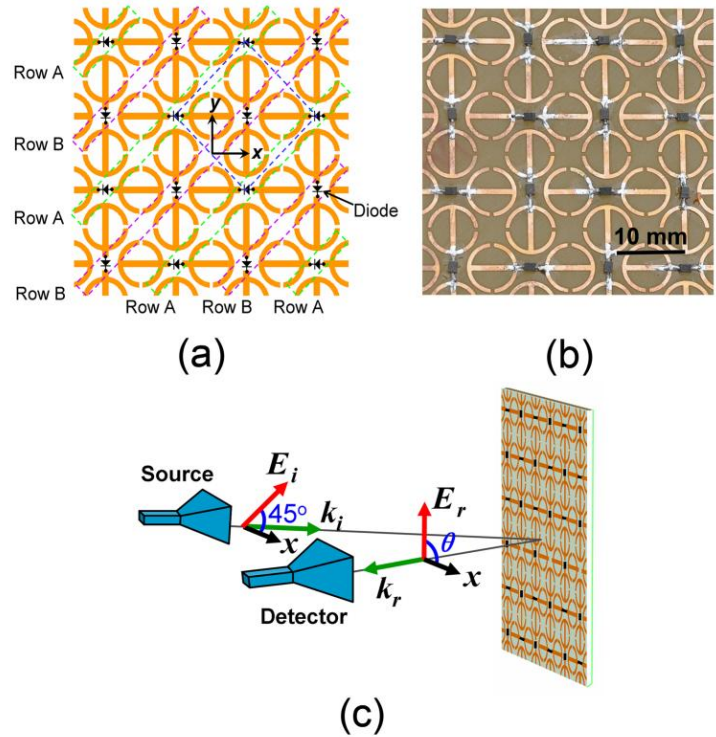

Fig. 1. (a) Schematic view of the tunable metamaterial reflector/absorber. The dashed blue square indicates the unit cell. (b) The fabricated sample. (c) Schematic view of the measurement arrangement.

In order to achieve polarization modulation, we firstly explore the tunable reflection characteristics of the metamaterial. We have numerically analyzed, physically fabricated and practically measured the proposed metamaterial structure. An EM simulator based on finite integration technique is used to optimize its various parameters. The metamaterial sample is designed on an FR4 dielectric substrate with the permittivity of 4.4, loss tangent of 0.02 and thickness of 2 mm. In simulation, the metallic structures on the top and bottom layers of the substrate are modeled as copper sheet with conductivity of $5.8 \times 10^7$ S/m and thickness of 17 µm. The ELC has inner and outer radius of 3 mm and 3.6 mm respectively, with two 0.5 mm gaps and a central strip of 0.8 mm in width. The microwave diode is modeled as a resistor in series with a capacitance at zero bias voltage and changes to a resistor in series with inductance at certain forward bias voltage. The parameters of the lumped elements are obtained by measuring the diode under different bias voltages at the working frequency using a RF impedance analyzer (Agilent E4991A).

Based on the optimized parameters, a sample (with outer dimension of $21 \times 21$ cm$^2$) is fabricated with print circuit board technique (PCB) (shown in Fig. 1(b)) and characterized by free space reflection measurement in a microwave anechoic chamber. A vector network analyzer (Agilent E8363C) and two linearly polarized horn antennas are used to transmit EM wave onto the sample and receive the reflected signal. Since the metamaterial has a full metallic film on the bottom layer so that the EM transmission is zero, and we only focus on its reflection characteristics. The incident angle is less than 5° from normal in the experiment, which can be regarded as a good approximation of normal incidence (Fig. 1(c)). The reflection measurement is calibrated by replacing the sample with an aluminum board of the same size as perfect reflector (unit reflection).

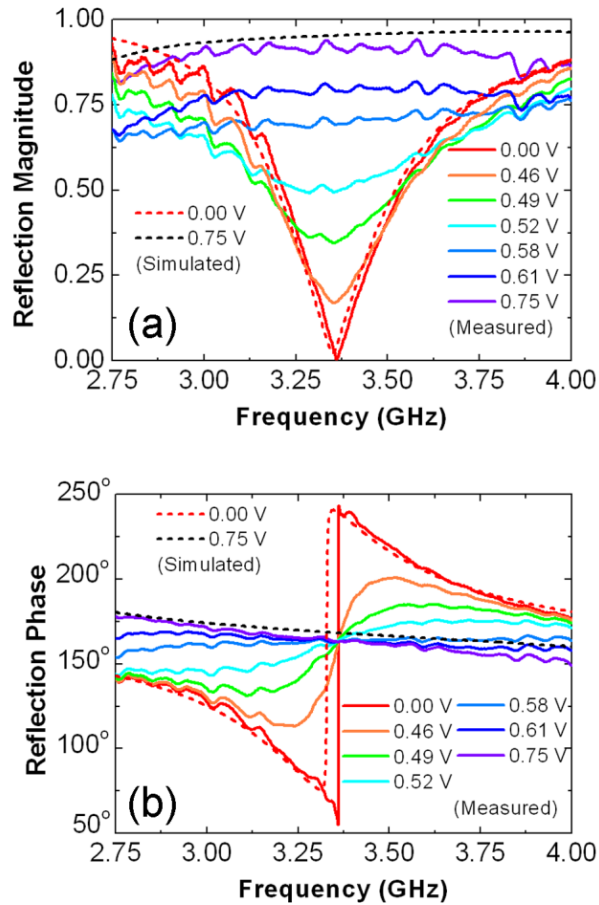

Fig. 2. Simulated and measured (a) magnitude and (b) phase of the reflection coefficient under normal incidence at various bias voltages of the diodes.

The simulated and measured magnitude and phase of the reflection are plotted in Fig. 2 under normal incidence. It can be observed from Fig. 2(a) that when the diodes are biased at zero voltages, the diodes are in OFF state so that all ELCs are coupled capacitively in series with a large diode resistance. The reflection magnitude undergoes a minimum (near zero) at 3.36 GHz, indicating a perfect absorption is achieved at this frequency. As increasing the forward bias voltage, the reflection magnitude increases gradually in this frequency band. This is because the diode impedance changes from capacitance in series with a large resistance to an inductance in series with a small resistance, leading to cancellation of the opposite charges on the ELC arches connected by the diode which gradually destroys the resonance of ELC and weakens the absorption. The maximum reflection magnitude (near unity) is finally achieved over the entire frequency band at the forward bias voltage of 0.75 V. The simulated reflection magnitude at the bias voltages of 0.00 V and 0.75 V are also plotted in Fig. 2(a) for comparison. They agree well with the corresponding measured curves verifying that the reflection magnitude of the metamaterial can be tuned continuously from zero to near unity by controlling the diode bias voltages.

Fig. 2(b) exhibits the reflection phase of the metamaterial under normal incidence. At zero bias voltage, strong EM resonance is established. The reflection phase firstly decreases with the increasing of the frequency, reaches a minimum value at the resonant frequency (3.36 GHz) and jumps to a maximum value abruptly, and then continues to decrease with the increasing of the frequency. When increasing the forward bias voltage, the resonance and absorption strength of the metamaterial is weakened, so that the phase jump at the resonance frequency is decreased and the phase changes slowly from minimum to maximum. A nearly flat phase response with a small negative gradient is finally achieved with forward bias voltage reaching 0.75 V since the resonance vanishes and the metamaterial behaves almost like a PEC reflector at this bias voltage. All the curves intersect each other at resonance frequency and each curve is nearly symmetric about the intersecting point. The simulated reflection phases at 0.00 V and 0.75 V bias voltages are in good agreement with the corresponding measured results except for a slight frequency shift, as depicted in Fig. 2(b).

In the case of oblique incidence, the reflection characteristics (both magnitude and phase) of the metamaterial remain unchanged within the incident angle of 20° [12].

## 3. Polarization modulation

As reported in [12], the ELCs in orthogonal orientations react to incident EM waves with the electric field polarized along $x$-axis and $y$-axis separately. Hence the reflection coefficient of the orthogonal polarized waves ($x$ or $y$ polarized waves) can be controlled independently through tuning the bias voltage applied on the diodes in row A and row B (Fig. 1(a)). This feature allows us to modulate the polarization state of the reflected waves flexibly.

Assume a normal incident wave with the electric field polarized along 45° azimuth angle with respect to $x$ axis. It can be decomposed into in phase $x$ polarized and $y$ polarized components with equal amplitude. Hence the electric field of the reflected wave can be denoted as $\boldsymbol{E}_r(\boldsymbol{r},t) = |\boldsymbol{E}_x|\cos(kz-\omega t)\hat{\boldsymbol{x}} + |\boldsymbol{E}_y|\cos(kz-\omega t+\varphi)\hat{\boldsymbol{y}}$, where $\varphi$, defined as $\varphi_y - \varphi_x$, is the phase difference between the $y$ and $x$ components. The values of $|\boldsymbol{E}_x|$ and $|\boldsymbol{E}_y|$ are proportional to the reflection magnitudes of the metamaterial for these orthogonal components as shown in Fig. 2(a), and $\varphi_x$, $\varphi_y$ take the values of the reflection phases of the metamaterial for the corresponding component as shown in Fig. 2(b). By tuning the bias voltage on the diodes in row A and row B, the amplitude and phase of the two orthogonal polarized components of the reflected wave can be adjusted independently so that the polarization state of the reflected wave can be controlled. The reflected wave becomes linearly polarized if $\sin\varphi = 0$, or left-handed (right-handed) elliptically polarized if $\sin\varphi > 0$ ($\sin\varphi < 0$). The axial ratio (AR) of the elliptical polarization state can be calculated by $20\log_{10}(|\boldsymbol{E}|_{\max}/|\boldsymbol{E}|_{\min})$. We also employ the polarization azimuth angle $\theta$ to describe the angle between the major polarization axis and the $x$-axis, which can be defined as $\theta = (1/2)\arctan\left[2|\boldsymbol{E}_x||\boldsymbol{E}_y|\cos\varphi/(|\boldsymbol{E}_x|^2 - |\boldsymbol{E}_y|^2)\right]$ [14].

Based on the measured reflection magnitude and phase shown in Fig. 2, we calculate the azimuth angle $\theta$ and axial ratio of the reflected waves at different bias voltages and working frequencies, under a 45° linearly polarized normal incident wave illuminating. The results are illustrated in Fig. 3. It is demonstrated in Fig. 3(a) that at the resonance frequency (3.36 GHz) of the metamaterial, the polarization azimuth angle of the reflected wave can be controlled by the bias voltages to change continuously within the range from 0° to 90°. Because $\varphi = 0$ at this frequency, the reflected wave is linearly polarized and the axial ratio is better than 35 dB as shown in Fig. 3(b).

The polarization of the reflected wave can be modulated by applying certain modulation bias voltages on the diodes in row A and row B. To give an example, we assume that two rectified sinusoidal modulation signals with $\pi/2$ phase delay are applied on the diodes in row A and B as demonstrated in Fig. 4. The EM wave reflected from the metamaterial can be modulated between $x$ and $y$ linear polarizations with the same frequency as the modulation bias signals.

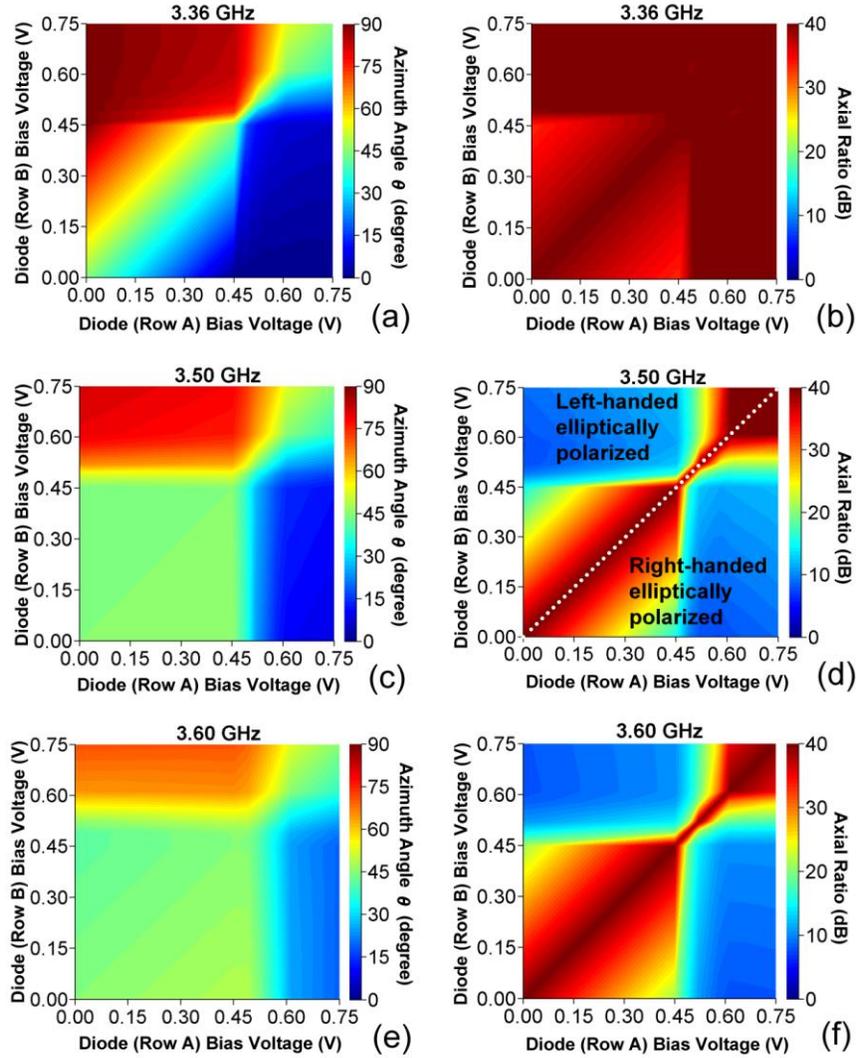

Fig. 3. Polarization state of the reflected wave under different bias voltages and working frequencies with a normal incident wave linearly polarized along 45° direction with respect to the $x$ axis. (a), (c), (e) denote polarization azimuth angle distributions, and (b), (d), (f) denote the corresponding axial ratio distributions.

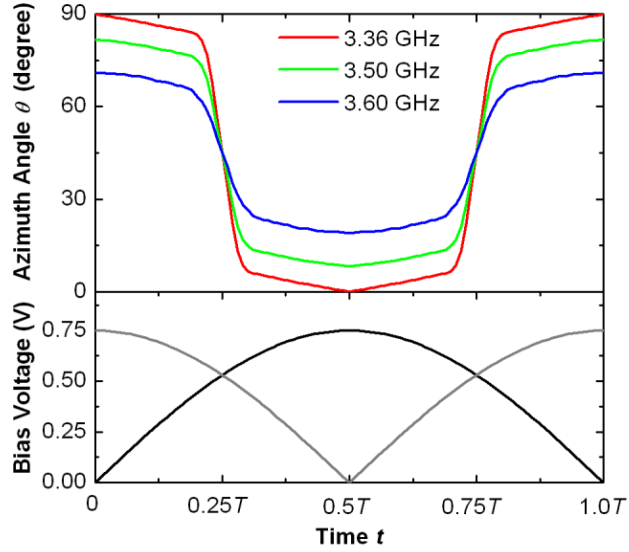

Fig. 4. The modulation of the azimuth angle θ of the major polarization axis with rectified sinusoidal modulation signals applied on the diodes in row A (black) and row B (grey) independently.

We also investigate the polarization modulation property when above the resonant frequency, as shown in Fig. 3(c) - 3(f). Similar properties can also be observed at the frequencies below the resonant frequency due to that the curves of reflection magnitude and phase are symmetric about the resonant frequency (Fig. 2). When off the resonant frequency, perfect absorption can not be achieved for either polarization component (Fig. 2(a)), so that the reflected wave contains both orthogonal polarization components. In addition, the reflection phases of $x$ and $y$ components are different if different bias voltages are applied on the diodes in row A and B (Fig. 2(b)), therefore the reflected wave becomes elliptically polarized with the superposition of the two orthogonal components.

From Fig. 3(c) - 3(d) and Fig. 4, it is found that at 3.50 GHz under the modulation bias voltages, the reflected wave changes from a left-handed elliptical polarization ($\sin \varphi > 0$) with an azimuth angle $\theta$ of 81° and AR of 8.8 dB (at $t = 0$) to a linear polarization with $\theta = 45°$ (at $t = 0.25T$), and then to a right-handed elliptical polarization ($\sin \varphi < 0$) with azimuth angle $\theta$ of 8° and AR of 8.8 dB (at $t = 0.5T$). When off the resonant frequency, the modulation range of the azimuth angle is reduced from the full range of 0° to 90°, which is due to that the phase variation range is less than 90° so that $\varphi$ is less than 90°. At 3.60 GHz, the modulation range of the azimuth angle is further shrunk to 19° to 71°, but AR is improved a little as shown in Fig. 3(e) - 3(f) and Fig. 4.

The presented polarization modulation scheme is based on the partial absorbing of certain polarized component by the metamaterial, therefore the amplitude of reflected wave is a little attenuated. However, the reflected power is no less than half of the incident power with the sinusoidal modulation signals.

In the practical applications, both linear and elliptical polarization modulation can be generated at different working frequencies with this metamaterial reflector/absorber. In addition to sinusoidal signals, other modulation signals such as square wave, or pulse signal can be used to achieve various kinds of polarization modulation flexibly. Although the sample is designed to function at microwave frequency, with the geometrical scalability, this concept can be applied at other frequency spectrums. The diodes can be replaced with semiconductor thin film structures integrated with the planar metamaterial structure as realized in [15] to operate at terahertz or optical range.

Recently, it is reported that metamaterial resonant structures loaded with nonlinear elements such as varactors can exhibit nonlinear phenomenon when the incident EM power is strong enough [16, 17]. The microwave diode used in the presented metamaterial structure is a nonlinear element, therefore theoretically there exists nonlinear response when the incident EM power is high enough, and the phase modulation property may be influenced. In our measurement, the transmitting antenna is placed about several meters away from the sample. We have measured the response of the metamaterial with different incident power (input power of the antenna) from −10 dBm to 20 dBm, but have not found any obvious nonlinear response. We believe that in this situation the incident EM power may not be strong enough to excite nonlinear effect of the metamaterial.

## 4. Conclusion

In conclusion, we have demonstrated a polarization modulation scheme based on the tunable metamaterial reflector/absorber which is constructed by coupling the ELC resonators in the metamaterial with microwave diodes. The metamaterial reacts independently to the orthogonal polarized incident waves with tunable reflection magnitude and phase through biasing the corresponding diodes. Illuminating the metamaterial with a $45^\circ$ linearly polarized wave, we are able to generate linear polarized waves in the reflection with polarization azimuth angle continuously tunable from $0^\circ$ to $90^\circ$ at the resonant frequency by controlling the bias voltages properly. When off the resonant frequency, elliptically polarized wave can be generated with tunable polarization azimuth angle of the major axis.

The proposed features allow us to electrically modulate the polarization state of EM waves flexibly by applying different modulation signals on the diodes. With geometrical scalability, this concept can be realized at terahertz or optical regions using standard lithography and replacing the diode with semiconductor thin film structure integrated with the resonant structures. We believe this polarization modulation method enables us to modulate EM wave polarizations flexibly, and is potentially useful in applications such as vibrational circular dichroism (VCD) spectroscopy, ellipsometry, and telecommunication devices [2, 10].

## Acknowledgments

This work is supported by the National Nature Science Foundation of China (Grant Nos. 6090320, 60990322, 60671002, and 60801001), and the National Basic Research Program of China (Grant No. 2004CB719800).